\documentclass[12pt]{article}
\usepackage{amssymb} 
\usepackage{graphicx}
\usepackage{amsmath}

\def\e{{\rm e}}

\def\d{\partial}
\def\l{\left(}
\def\r{\right)}

\newcommand{\be}{\begin{equation}}
\newcommand{\ee}{\end{equation}}
\newcommand{\ba}{\begin{align}}
\newcommand{\ea}{\end{align}}
\newcommand{\bg}{\begin{gather}}
\newcommand{\eg}{\end{gather}}
\newcommand{\bseq}{\begin{subequations}}
\newcommand{\eseq}{\end{subequations}}

\renewcommand{\ln}{\mathop{\rm ln}\nolimits}

\renewcommand{\Re}{\mathop{\rm Re}\nolimits}

\makeatletter
\def\gsim{\compoundrel>\over\sim}
\def\lsim{\compoundrel<\over\sim}
\def\compoundrel#1\over#2{\mathpalette\compoundreL{{#1}\over{#2}}}
\def\compoundreL#1#2{\compoundREL#1#2}
\def\compoundREL#1#2\over#3{\mathrel
         {\vcenter{\hbox{$\m@th\buildrel{#1#2}\over{#1#3}$}}}}
\makeatother

\begin{document}

\begin{center}
  {\Large\bf Ultra-large distance modification of gravity from
  Lorentz symmetry breaking at the Planck scale}\\
\medskip
D.S.~Gorbunov, S.M.~Sibiryakov 
\\
\medskip
{\small
Institute for Nuclear Research of
         the Russian Academy of Sciences,\\  60th October Anniversary
  prospect, 7a, 117312 Moscow, Russia
  }
\end{center}

\begin{abstract}
We present an extension of the Randall--Sundrum model in which, due to
spontaneous Lorentz symmetry breaking, graviton mixes with bulk
vector fields and becomes quasilocalized. The masses of KK modes
comprising the four-dimensional graviton are naturally exponentially
small. This allows to push the Lorentz breaking scale to as high as a
few tenth of the Planck mass. The model does not contain ghosts or
tachyons and does not exhibit the van Dam--Veltman--Zakharov
discontinuity. The gravitational attraction between static point
masses becomes gradually weaker with increasing of separation and gets 
replaced by repulsion (antigravity) at exponentially large
distances.
\end{abstract}


\section{Introduction}
\label{intro}
The discovery of cosmological acceleration \cite{Riess:1998cb}
indicates the necessity for modification of the standard cosmological
paradigm.  One way to explain the accelerated expansion of the
Universe is to attribute it to the effect of some unconventional
matter such as the cosmological constant or quintessence. In
this approach one encounters, however, the difficulty of explaining
the small value of the associated energy scale in comparison with
other fundamental scales, such as the Planck mass or electroweak symmetry
breaking scale.  It is reasonable to consider an alternative possibility
that the cosmological constant is exactly zero, but its effect is
mimicked by modifications of gravitational laws at distance scale of the
order of the present cosmological horizon.

A natural framework for modifying gravity at ultra-large distances is
provided by models with infinite extra spatial dimensions.  A number
of brane-world models have been proposed
\cite{Gregory:2000jc,Dvali:2000hr} where the
four-dimensional graviton is a superposition of bulk KK modes, and
thus, is not localized on the brane. Rather, it is quasilocalized,
having a non-zero width $\Gamma$ (and, possibly, mass $m$) with
respect to decay into extra dimensions. In these models gravity is
four-dimensional at distances smaller than $r_c\sim
\min\{m^{-1},\Gamma^{-1}\}$ and becomes multi-dimensional at larger
scales. The attractive feature of the models with extra-dimensions is
that the tiny values of $m, \Gamma$ are generated without strong
fine-tuning of parameters.

However, Lorentz invariant extra dimensional models with infrared
modifications of gravity face the problems
\cite{Pilo:2000et,Luty:2003vm,Rubakov:2003zb} akin to those of four-dimensional
massive gravity. Either these theories contain ghosts, fields with the
wrong sign of kinetic term, or the propagator of graviton exhibits
the van Dam--Veltman--Zakharov (vDVZ) discontinuity
\cite{vanDam:1970vg} originating from a scalar degree of
freedom that does not decouple in the massless limit. At the classical
level the discontinuity might be cured by nonlinear effects
\cite{Vainshtein:1972sx,Nicolis:2004qq},
but at the quantum level the presence of the additional scalar leads
to strong coupling
\cite{Arkani-Hamed:2002sp,Luty:2003vm,Rubakov:2003zb} 
at the energy scale at
most of order $(m^2M_{Pl})^{1/3}$.
If $m$ is comparable to the present 
Hubble parameter of the Universe, the theory is in strong coupling
regime, and thus looses its predictivity, at distances smaller than
1000 km.

There are indications that these difficulties might be resolved in the
Dvali--Gabadadze--Porrati (DGP) \cite{Dvali:2000hr} model. In
particular, it was argued that the scale of strong coupling can be
pushed to higher energies in nontrivial classical backgrounds,
provided the structure of counterterms in the model is of a special
form \cite{Nicolis:2004qq}. However, it is still unclear whether the
counterterms in the DGP model actually meet the requirements of
Ref.~\cite{Nicolis:2004qq}. Probably, the most disappointing fact
about the DGP model is that the branch of the 
cosmological solutions in this theory,
which exhibits cosmic acceleration without introduction of the
cosmological constant, is plagued by ghost-like
instabilities
\cite{Luty:2003vm,Nicolis:2004qq}.

In this paper we propose a brane-world model with infrared
modification of gravity and violation of 4-dimensional 
Lorentz-invariance. In
our model graviton becomes quasilocalized due to mixing with vector
fields which freely propagate in the bulk. This kind of mixing becomes
possible once the Lorentz symmetry is spontaneously broken by
condensates of the 
vector fields.

Our approach builds up on the ideas put forward recently in the
four-dimensional framework. It was suggested that the problems of
massive gravity could be resolved in models incorporating violation of
Lorentz invariance
\cite{Arkani-Hamed:2003uy,Rubakov:2004eb,Dubovsky:2004sg}.  Models
with explicit Lorentz violation in the gravitational sector were
considered in
Refs.~\cite{Arkani-Hamed:2003uy,Rubakov:2004eb,Dubovsky:2004sg,
Dubovsky:2004ud} and it was shown that they are equivalent to a class
of models with gravity coupled to scalar fields with unusual kinetic
terms.  In the latter formulation the Lorentz symmetry is broken
spontaneously by coordinate dependent scalar condensates. Because of
nonlinear structure of the scalar kinetic term in this kind of models,
they possess a cutoff scale $\Lambda_{cutoff}=(m M_{Pl})^{1/2}$, and
can be considered only as low energy effective $\sigma$-models. If the
graviton mass $m$ is of the order of the present Hubble parameter,
then $\Lambda_{cutoff}\approx (0.01\mathrm{mm})^{-1}$, which is
phenomenologically acceptable, but still unnaturally small in
comparison with other fundamental scales. It is unclear whether these
$\sigma$-models can be extended to complete models making sense above
the scale $\Lambda_{cutoff}$.

As a scenario for such a complete model one could envisage a
gravitational analog of the Higgs mechanism triggered by spontaneous
Lorentz symmetry breaking.  In
Refs.~\cite{Kostelecky:1989jp,Clayton:1998hv,Jacobson:2001yj,Gripaios:2004ms,
Carroll:2004ai} 
it was proposed to use
coordinate independent vector (in general, tensor)
condensates for this purpose. 
However, in four-dimensional theories, graviton
does not acquire mass within this approach: 
no gap in the dispersion relation of graviton
appears \cite{Kostelecky:1989jw,LibRub}. Moreover, the effect of
vector condensates on cosmology amounts to nothing but the change of
the gravitational constant in the cosmological equations
\cite{Carroll:2004ai,LibRub}, and no cosmological acceleration is
produced.

The brane-world approach opens up a new way to circumvent the problems
encountered in four dimensions. In this paper we make the first step
along this line: in Sec.~2 we present a model in which
graviton becomes quasilocalized due to spontaneous Lorentz symmetry
breaking induced by non-zero VEVs of bulk vector fields, and analyze
in the subsequent sections the spectrum of linear perturbations about
four-dimensionally flat background.  In Sec.~3 we analyze the tensor
perturbations and find that the characteristic mass $m_c$ of KK modes
comprising four-dimensional graviton is naturally exponentially
small, which allows to take the Lorentz-breaking scale as high as a
few tenth of the Planck mass.  In Sec.~4 and Sec.~5 we study vector
and scalar perturbations and show that the model contains neither 
tachyons, nor 
ghosts, and does not exhibit the vDVZ discontinuity. We calculate long
distance modification of the Newton law and find quite unexpectedly
that attraction of point masses changes to repulsion at distances of
order $1/m_c$. Thus, the model provides a realization of antigravity
at ultra-large distances. This looks promising in view of the
possibility to obtain cosmological acceleration at late times. On the
other hand, modification of gravitational field of static sources in
the model faces severe phenomenological constraints.  We discuss
possible ways to get around these constraints in the concluding
Sec.~6.

A number of important issues are left beyond the scope of the present
paper. These are, e.g., cosmology of the proposed model and structure
of quantum
corrections. We leave them for future investigation.


\section{The model}
\label{model}

We consider a five-dimensional setup with positive tension brane and
negative cosmological constant in the bulk --- the
Randall--Sundrum (RS) model \cite{Randall:1999vf}. 
We add to this model
three bulk vector fields $A_{M}^a$, $a=1,2,3$,
with quartic potential localized on the brane. 
The action is taken in the following form,
\be
\begin{split}
S=&\int d^5x\sqrt{g}\bigg(-\frac{R}{16\pi G_5}-\Lambda -
\frac{1}{4}F^a_{MN}F^{a\,MN}\bigg)\\
&+\int d^4x\sqrt{-\bar{g}}\left(-\sigma-
\frac{\varkappa_1^2}{2}
\big(\bar{g}^{\mu\nu}A_{\mu}^aA^{b}_{\nu}+v^2\delta^{ab}\big)^2-
\frac{\varkappa_2^2}{2}
\left(\bar{g}^{\mu\nu}A_{\mu}^aA^{b}_{\nu}-
\frac{1}{3}\delta^{ab}\bar{g}^{\mu\nu}A_{\mu}^cA^{c}_{\nu}\right)^2
\right)\;,
\end{split}
\label{action}
\ee
where $\bar{g}_{\mu\nu}$ is the induced metric on the brane, and
\[
F_{MN}^a=\d_{M}A_N^a-\d_N A_M^a\;.
\]
The capital Latin indices $M,N$ take values $0,1,2,3,5$, while the
Greek indices $\mu,\nu,\ldots$ run from $0$ to $3$.  The action
(\ref{action}) is invariant under global $SO(3)$ symmetry with vector
fields $A_\mu^a$ belonging to fundamental representation. It is
straightforward to check that the potential in the second line
of Eq.~(\ref{action}) is generic quartic potential invariant under
this group.  Note that the second term in the potential depends only
on the traceless part of the matrix
$\bar{g}^{\mu\nu}A_{\mu}^aA^{b}_{\nu}$.  The parameters $v^2$,
$\varkappa_1^2$, $\varkappa_2^2$ are assumed to be positive.  In
addition to the global $SO(3)$ symmetry, the bulk part of the action is
invariant under $U(1)\times U(1)\times U(1)$ gauge transformations,
\[
A^a_M\mapsto A^a_M+\d_M\alpha^a\;.
\]
This Abelian gauge symmetry is broken explicitly at the brane.

The model has a static solution with AdS metric in
the bulk and constant values of the vector fields,
\bseq
\label{stat*}
\begin{align}
\label{stat1}
&ds^2=-dz^2+\e^{-2k|z|}\eta_{\mu\nu}dx^{\mu}dx^{\nu}\;,\\
\label{stat2}
&A_5^a=A_0^a=0~,~~A^a_i=v\delta^a_i~,~~i=1,2,3\;,
\end{align}
\eseq where $k=\sqrt{-4\pi G_5\Lambda/3}$, and the usual fine-tuning
$\sigma=-\Lambda/k$ is assumed; the signature of the Minkowski metric
$\eta_{\mu\nu}$ is $(+,-,-,-)$. The VEVs (\ref{stat2}) of the vector
fields break $SO(3)\times$Lorentz symmetry down to $SO(3)$ of spatial
rotations accompanied by simultaneous rotations in the internal
space. The similar pattern of Lorentz symmetry breaking was considered
in the four-dimensional context in Refs.~\cite{Bento:1992wy,LibRub}.
As we will see, this spontaneous symmetry breaking provides
mixing between the graviton zero mode, present in the pure RS case,
and the continuum spectrum of KK modes of vector fields. This mixing
results in quasilocalization of the graviton.

Let us study the linearized perturbations above the background
(\ref{stat*}). Imposing the gauge conditions $g_{5\mu}=0$, $g_{55}=-1$
on the metric, one writes the following decomposition, 
\begin{align*}
&ds^2=-dz^2+(\e^{-2k|z|}\eta_{\mu\nu}+h_{\mu\nu}(x,z))dx^{\mu}dx^{\nu}\;,\\
&A^a_M=v\delta^a_M+a^a_M(x,z)\;.
\end{align*}
The analysis is simplified by the fact that the energy-momentum tensor of the
vector fields in the bulk vanishes to the linear order in
perturbations, allowing for fixing the transverse traceless gauge in
the bulk \cite{Randall:1999vf,Garriga:1999yh}, 
\be
\label{trtr} 
h_{\mu}^{\mu}=h_{\nu,\mu}^{\mu}=0\;.  \ee Here and throughout the
paper indices $\mu,\nu,\dots$ are raised (lowered) using the metric
$\eta^{\mu\nu}$ ($\eta_{\mu\nu}$), and comma denotes
derivative. In this gauge the bulk equations for the metric take a
fairly simple form, \be
\label{hbulkeq}
\frac{1}{2}h''_{\mu\nu}-2k^2h_{\mu\nu}-
\frac{1}{2[u(z)]^2}h_{\mu\nu,\lambda}^{~~,\lambda}=0\;,
\ee  
where $u(z)=\e^{-k|z|}$, and prime denotes derivative with respect
to $z$. As to the vector fields, the bulk equations read
\be
\label{abuleq}
-{a_{\nu}^a}''+2k{a^a_{\nu}}'+\frac{1}{u^2}a^{a\;,\mu}_{\nu\;,\mu}=0\;.
\ee 
In deriving this equation we made use of the $[U(1)]^3$ gauge
invariance of the bulk action to impose the conditions
\be
\label{Agauge}
a^a_5=0\;, ~~a^{a\,\mu}_{~~~,\mu}=0\; 
\ee 
on the vector fields in the
bulk. Imposing the two conditions
\eqref{Agauge} simultaneously 
is possible, as they are compatible on-shell.  The
coordinate frame where the conditions \eqref{trtr} and \eqref{Agauge} 
are satisfied will be referred to below as the bulk frame. 

The boundary conditions on the brane are most easily formulated in the
Gauss normal (GN) reference frame. In this frame the brane
is fixed at $\bar z=0$ (the quantities with the bar refer to
the GN frame), and we assume $Z_2$ symmetry across the brane. 
Then, 
for general energy-momentum tensor $T_{\mu\nu}$ on the
brane, the boundary conditions for the metrics at $\bar z=+0$  
read \cite{Garriga:1999yh},
\be
\label{boundmunu}
\bar h'_{\mu\nu}+2k\bar h_{\mu\nu}=
8\pi
G_5\left(T_{\mu\nu}-\frac{1}{3}\eta_{\mu\nu}T^{\lambda}_{\lambda}\right)\;. 
\ee
The 
linearized energy-momentum tensor of the vector fields 
comes from the Lorentz violating term in the 
quadratic action arising from the brane potential in
Eq.~(\ref{action}). To the quadratic order, the latter term 
has the form,
\be
\label{LV}
\begin{split}
S^{(2)}_{LV}=-\int d^4x&\bigg[
\frac{\varkappa_1^2v^4}{2}
\left(\bar h_{ab}+\frac{1}{v}(\bar a^a_b+\bar a^b_a)\right)^2\\
&+\frac{\varkappa_2^2v^4}{2}
\left(\bar{h}_{ab}+\frac{1}{v}(\bar{a}^a_b+\bar{a}^b_a)-
\frac{1}{3}\delta_{ab}\left(\bar{h}_{cc}+
\frac{2}{v}\bar{a}^c_c\right)\right)^2\bigg]\;.
\end{split}
\ee
From this expression one obtains,
\bseq
\label{Tmunu}
\begin{align}
\label{T000a}
&T^V_{00}=T^V_{0a}=0\;,\\
&T^V_{ab}=2\varkappa^2v^4\left(\bar h_{ab}+
\frac{1}{v}(\bar a^a_b+\bar a^b_a)\right)
-\frac{2\varkappa_2^2v^4}{3}\delta_{ab}\left(\bar{h}_{kk}+
\frac{2}{v}\bar{a}^k_k\right)\;,
\end{align}
\eseq
where we introduced $\varkappa^2=\varkappa_1^2+\varkappa_2^2$.
Note that the energy-momentum tensor (\ref{Tmunu}) of the vector fields  
violates the weak energy condition. 
We will show that this property does not lead to
instabilities. At the same time it is crucial for
antigravity emerging at ultra-large distances.

For the vector fields, the $Z_2$ symmetry and continuity demand that the
5th component  
vanishes on the brane, $\bar a^a_5(\bar z=0)=0$. The sources for the
other components can be read off from (\ref{LV}), resulting in  
the following boundary conditions at $\bar z=+0$,
\bseq
\label{abarbound*}
\begin{align}
\label{abarbaound1}
& \bar a^{a'}_0=0\;,\\
\label{abarbound2}
&\bar a^{a'}_i=\varkappa^2v^3\left(\bar h_{ai}+\frac{1}{v}(\bar
a^a_i+\bar a^i_a)\right)
-\frac{\varkappa_2^2v^3}{3}\delta_{ai}
\left(\bar{h}_{kk}+\frac{2}{v}\bar{a}^k_k\right)\;. 
\end{align} 
\eseq 

The metrics in the bulk frame and GN frame are related by a gauge
transformation. The form of the latter is restricted by the condition
$\bar h_{55}=\bar h_{5\mu}=0$. One obtains 
\be
\label{hGN}
\bar h_{\mu\nu}=h_{\mu\nu}+u^2(\varepsilon_{\mu,\nu}+\varepsilon_{\nu,\mu})
+\frac{1}{k}\varepsilon_{,\mu\nu}-2ku^2\eta_{\mu\nu}\varepsilon\;,
\ee  
where the functions 
$\varepsilon$, $\varepsilon_{\mu}$ depend only on $x$. 
This transformation corresponds to the 
following change of coordinates,
\[
z=\bar z+\varepsilon\;,~~~
x^{\mu}=\bar x^{\mu}+\frac{1}{2ku^2}\varepsilon^{,\mu}+\varepsilon^{\mu}\;.
\]
Note that in the bulk frame the brane is displaced from the origin along
the $z$-direction, its position being described by
the brane bending $\varepsilon$. 
The relation between the components of the 
vector fields in the two frames has the form
\begin{align*}
\bar a^a_5&=a^a_5-\frac{v}{u^2}\varepsilon_{,a}+\d_5\alpha^a\;,\\
\bar a^a_{\mu}&=a^a_{\mu}-\frac{v}{2ku^2}\varepsilon_{,a\mu}
+v\varepsilon^a_{,\mu}+\alpha^a_{,\mu}\;,
\end{align*}
where we allowed for the possibility of a $[U(1)]^3$ gauge
transformation parametrized by the functions $\alpha^a(z,x)$.
Imposing the gauge $\bar a^a_5=0$, which is consistent with the
boundary conditions on the brane, we obtain,
\be
\label{aGN}
\bar a^a_{\mu}=a^a_{\mu}+v\varepsilon^a_{,\mu}+\beta^a_{,\mu}\;,
\ee
where the functions $\beta^a$ depend only on $x$. 
The set of equations (\ref{boundmunu}), (\ref{Tmunu})--(\ref{hGN}) and
(\ref{aGN}) results in the following boundary conditions,
\bseq
\label{hbound*}
\begin{align}
\label{hbound00}
h_{00}'+2kh_{00}+2\varepsilon_{,00}=&
\frac{\lambda_1}{3}\left(h_{aa}+\frac{2}{v}a^a_a
+\frac{1}{k}\varepsilon_{,aa}+6k\varepsilon+\frac{2}{v}\beta^a_{~,a}\right)
\;,\\
\label{hbound0i}
h'_{0i}+2kh_{0i}+2\varepsilon_{,0i}=0&\;,\\
\notag
h_{ij}'+2kh_{ij}+2\varepsilon_{,ij}=\lambda&\bigg\{h_{ij}
+\frac{1}{v}(a^i_j+a^j_i)+\frac{1}{k}\varepsilon_{,ij}+
\frac{1}{v}(\beta^i_{,j}+\beta^j_{,i})\\
&-\frac{1}{3}\delta_{ij}
\left(h_{kk}+\frac{2}{v}a^k_k+\frac{1}{k}\varepsilon_{,kk}
+\frac{2}{v}\beta^k_{~,k}\right)\bigg\}\;,
\label{hboundij}\\
\label{abounda0}
{a^a_0}'=0\;,~~~~~~~~~~~~~~~~~~&\\
\notag
{a^a_i}'=\varkappa^2v^3\bigg\{h_{ai}+\frac{1}{v}(a^a_i&+a^i_a)+
\frac{1}{k}\varepsilon_{,ai}+\frac{1}{v}(\beta^a_{,i}+\beta^i_{,a})
\bigg\}+2\varkappa_1^2v^3k\delta_{ai}\varepsilon\\
&-\frac{\varkappa_2^2v^3}{3}\delta_{ai}
\left(h_{kk}+\frac{2}{v}a^k_k+\frac{1}{k}\varepsilon_{,kk}
+\frac{2}{v}\beta^k_{,k}\right)\;;
\label{aboundai}
\end{align}
\eseq
here we introduced the notations 
\[
\lambda_1=16\pi G_5\varkappa_1^2v^4 
\;,~~~~\lambda=16\pi G_5\varkappa^2v^4\;.
\]

In what follows it is convenient to work in 
the 4-dimensional Fourier representation,
\[ 
h_{\mu\nu}, a^a_{\mu} \propto \e^{-ip_0x_0+i{\bf px}}\;,
\]
and decompose the perturbations into scalar, vector and tensor modes
with respect to rotations around the three-momentum ${\bf p}$.  To
this end one introduces a three-dimensional orthogonal basis
consisting of ${\bf p}$ and two unit vectors, ${\bf e}^{(\alpha)}$,
$\alpha =1,2$. The latter are used to construct a pair of
transverse traceless symmetric tensors,
\[
d^{(1)}_{ij}=\frac{1}{\sqrt{2}}(e_i^{(1)}e_j^{(2)}+e_j^{(1)}e_i^{(2)})~,~~~
d^{(2)}_{ij}=\frac{1}{\sqrt{2}}(e_i^{(1)}e_j^{(1)}-e_i^{(2)}e_j^{(2)})\;,
\]
and a transverse  antisymmetric tensor,
\[
f_{ij}=\frac{1}{\sqrt{2}}(e_i^{(1)}e_j^{(2)}-e_j^{(1)}e_i^{(2)})\;.
\]
The tensorial decomposition reads,
\bseq
\label{decomp*}
\begin{align}
\label{decomph00}
h_{00}=&\phi_1\;,\\
\label{decomph0i}
h_{0i}=&p_i\phi_2+e_i^{(\alpha)}\psi_{1\alpha}\;,\\
\label{decomphij}
h_{ij}=&\left(\delta_{ij}-\frac{p_ip_j}{{\bf p}^2}\right)\phi_3
+p_ip_j\phi_4+\left(p_ie_j^{(\alpha)}+p_je_i^{(\alpha)}\right)\psi_{2\alpha}
+d_{ij}^{(\alpha)}\chi_{1\alpha}\;,\\
\label{decompaa0}
a^a_0=&p_a\phi_5+e_a^{(\alpha)}\psi_{3\alpha}\;,\\
\notag
a^a_i=&\left(\delta_{ia}-\frac{p_ip_a}{{\bf
p}^2}\right)\phi_6+p_ip_a\phi_7
+\left(p_ie_a^{(\alpha)}+p_ae_i^{(\alpha)}\right)\psi_{4\alpha}+
\left(p_ie_a^{(\alpha)}-p_ae_i^{(\alpha)}\right)\psi_{5\alpha}\\
\label{decompaai}
&+d_{ia}^{(\alpha)}\chi_{2\alpha}+f_{ia}\xi\;,\\
\label{decompbetaa}
\beta^a=&p_a\phi_8+e_a^{(\alpha)}\psi_{6\alpha}\;,
\end{align}
\eseq
with ${\bf p}=(p_1,p_2,p_3)$. 
There are two fields $\chi$ in the symmetric tensor sector; one field
$\xi$ in the sector of antisymmetric tensors; six fields $\psi$ in the
vector sector, and eight fields $\phi$ in the scalar sector. Note that
all fields in (\ref{decomp*}), except for $\phi_8$ and $\psi_6$, are
functions of the fifth coordinate $z$.


\section{Tensor modes}
\label{temsors}

Let us study the spectrum of tensor perturbations. To begin with, 
one notices
that the antisymmetric field $\xi$ completely decouples. 
It satisfies Eq.~(\ref{abuleq}) with the free boundary condition
$\xi'(z=0)=0$, and forms a continuum spectrum of completely 
delocalized modes similar to the modes of a free vector field in
RS background \cite{Dubovsky:2000av}. We do not
consider these modes below.

For the symmetric tensors one obtains from Eqs.~(\ref{hbulkeq}), 
(\ref{abuleq}) the
following equations in the bulk,
\bseq
\label{tens*}
\begin{align}
\label{tens1}
&\frac{1}{2}\chi_1''-2k^2\chi_1+\frac{p^2}{2u^2}\chi_1=0\;,\\
\label{tens2}
&-\chi_2''+2k\chi_2'-\frac{p^2}{u^2}\chi_2=0~,~~~~~z>0\;,
\end{align}
\eseq
where $p^2\equiv p_0^2-{\bf p}^2$, and 
polarization index $\alpha$ is omitted to simplify notations.
The boundary conditions are read off from Eqs.~(\ref{hboundij}),
(\ref{aboundai}): 
\bseq
\label{btens*}
\begin{align}
\label{btens1}
&\chi_1'+2k\chi_1=\lambda\left(\chi_1+\frac{2}{v}\chi_2\right)\;,\\
\label{btens2}
&\chi_2'=\varkappa^2v^3\left(\chi_1+\frac{2}{v}\chi_2\right)~,~~~~~~z=+0\;.
\end{align}
\eseq
The system (\ref{tens*}), (\ref{btens*}) 
can be rewritten in the form of an eigenvalue
problem,
\bseq
\label{eitens*}
\begin{align}
\label{eitens1}
-u^2(\chi_1''-4k^2\chi_1)-
\big[(4k-2\lambda)\chi_1-2\mu\hat{\chi}_2\big]\delta(z)=&m^2\chi_1\;,\\
\label{eitens2}
-u^2(\hat{\chi}_2''-2k\hat{\chi}_2')+
\big[2\mu\chi_1+2\nu\hat{\chi}_2\big]\delta(z)=&m^2\hat{\chi}_2\;,
\end{align}
\eseq where we denoted $m^2=p^2$, $\nu=2\varkappa^2v^2$,
$\mu=\sqrt{\lambda\nu}$ and introduced $\hat{\chi}_2=\sqrt{32\pi
G_5}\chi_2$ to make the operator Hermitian with the scalar
product 
\be
\label{norm}
(\eta,\chi)=\int_{-\infty}^{+\infty} dz \left(\frac{\eta^*_1\chi_1}{u^2}+
\hat{\eta}^*_2\hat{\chi}_2\right)\;.
\ee

First, let us make sure that there are no tachyonic modes. 
The operator (\ref{eitens*}) is the sum of
the diagonal operator appearing in the pure RS case and a
positive semi-definite operator
\[
\Delta=\begin{pmatrix}
2\lambda & 2\mu\\
2\mu & 2\nu
\end{pmatrix}
\delta(z)\;.
\]
Thus, the absence of tachyons in our case is ensured by their absence in
the RS model.

Next, there is no normalizable zero mode. Indeed, if 
$m^2=0$, then $\chi_1=A_1\e^{-2k|z|}$, $\hat{\chi}_2=A_2=const$ (other
solutions  to
Eqs.~(\ref{tens*}) are exponentially growing). From Eqs.~(\ref{btens*})
one obtains that $2\lambda A_1+2\mu A_2=0$. Thus, $A_2\neq 0$ and 
the zero mode is not normalizable with the scalar product (\ref{norm}). 

We now turn to the modes with $m^2\geq 0$. These modes belong to 
continuum spectrum. Introducing the conformal
coordinate $\zeta=\e^{k|z|}/k$
one writes the general solution of the bulk equations (\ref{tens*}) 
as a combination of Bessel functions, 
\bseq
\label{bulksol*}
\begin{align}
\label{bulksol1}
&\chi_1=A_1J_2(m\zeta)+B_1N_2(m\zeta)\;,\\
\label{bulksol2}
&\hat{\chi}_2=m\zeta\bigl[A_2J_1(m\zeta)+B_2N_1(m\zeta)\bigr]\;,
\end{align}
\eseq 
with $A_1, B_1, A_2, B_2$ being complex numbers. For each value of the
mass $m$ there are two eigenvectors $\chi^{(r)}_m$, $r=1,2$, of 
Eq.~(\ref{eitens*}), which are normalized as follows,
\be
\label{normkl}
\Big(\chi_m^{(r)},\chi_{m'}^{(s)}\Big)=\delta^{rs}\delta(m-m')\;.
\ee
The normalization factor is determined by the asymptotics of the modes at large
$\zeta$. Making use of the asymptotics of the expressions
(\ref{bulksol*}) one translates (\ref{normkl}) into 
the relation between the coefficients
$A,B$,
\be
\label{normAB}
\frac{2k}{m}
\Big(\big(A_{1m}^{(r)}\big)^*A_{1m}^{(s)}+
\big(B_{1m}^{(r)}\big)^*B_{1m}^{(s)}\Big)+
\frac{2m}{k}\Big(\big(A_{2m}^{(r)}\big)^*A_{2m}^{(s)}+
\big(B_{2m}^{(r)}\big)^*B_{2m}^{(s)}\Big)=\delta^{rs}\;.
\ee
Let us turn to the boundary equations (\ref{btens*}). 
In terms of the coefficients
$A,B$ they take the form,
\bseq
\label{tjunc*}
\begin{align}
\notag
A_1\left(J_1\left(\frac{m}{k}\right)-
\frac{\lambda}{m}J_2\left(\frac{m}{k}\right)\right)
&+B_1\left(N_1\left(\frac{m}{k}\right)-
\frac{\lambda}{m}N_2\left(\frac{m}{k}\right)\right)\\
\label{tjunc1}
&-A_2\frac{\mu}{k}J_1\left(\frac{m}{k}\right)
-B_2\frac{\mu}{k}N_1\left(\frac{m}{k}\right)=0\;,\\
\notag
-A_1\frac{\mu}{m}J_2\left(\frac{m}{k}\right)
-B_1\frac{\mu}{m}N_2\left(\frac{m}{k}\right)
&+A_2\left(\frac{m}{k}J_0\left(\frac{m}{k}\right)-
\frac{\nu}{k}J_1\left(\frac{m}{k}\right)\right)\\
\label{tjunc2}
&+B_2\left(\frac{m}{k}N_0\left(\frac{m}{k}\right)-
\frac{\nu}{k}N_1\left(\frac{m}{k}\right)\right)=0\;.
\end{align}
\eseq
It is convenient to choose the first eigenvector of the operator
(\ref{eitens*}) to satisfy the relation
\[
\lambda\chi_1^{(1)}+\mu\hat{\chi}_2^{(1)}\big|_{z=0}=0\;.
\] 
Then Eqs.~(\ref{tjunc*}) yield
\begin{align*}
&A_1^{(1)}=C^{(1)}N_1\left(\frac{m}{k}\right)~,&
&B_1^{(1)}=-C^{(1)}J_1\left(\frac{m}{k}\right)\;,\\
&A_2^{(1)}=-C^{(1)}\frac{\lambda k}{\mu
m}N_0\left(\frac{m}{k}\right)~,&
&B_2^{(1)}=C^{(1)}\frac{\lambda k}{\mu
m}J_0\left(\frac{m}{k}\right)\;,
\end{align*}
where the normalization constant $C^{(1)}$ is fixed by
Eq.~(\ref{normAB}). At $m\ll k$ it has the form,
\[
C^{(1)}=\pi\left(\frac{m}{2k}\right)^{3/2}\;,
\]
implying the following amplitude of the graviton field on the brane
\[
\chi_1^{(1)}\big|_{z=0}=\left(\frac{m}{2k}\right)^{1/2}\;.
\]
These modes are analogous to the continuum graviton spectrum in the RS
case. They are completely delocalized and provide corrections to
the four-dimensional Einstein gravity only at short distances $r\sim 1/k$.

The orthogonal mode is fixed by Eq.~(\ref{normAB}). In the general
case the corresponding expressions are rather complicated. A
considerable simplification occurs, if one assumes the
hierarchy between the parameters,
$\lambda \ll \mu\ll\nu\ll k\sim M_5$, where $M_5=(16\pi G_5)^{-1/3}$
is the five-dimensional Planck mass. 
Since the physics at large distances is governed by light modes we are
interested in the region of masses $m^2\ll\lambda k$. One obtains,
\begin{align*}
&A_1^{(2)}=\frac{2\mu}{\nu}\left(\frac{m}{2k}\right)^{3/2}
\left(\ln{\frac{m}{2k}}+\boldsymbol{C}\right)~,&
&B_1^{(2)}=\frac{\pi\mu}{\nu}\left(\frac{m}{2k}\right)^{3/2}
\frac{1}{1-\frac{2\lambda}{\nu}\ln{\frac{m}{k}}}\;,\\
&A_2^{(2)}=-\frac{1}{2}\left(\frac{2k}{m}\right)^{1/2}~,&
&B_2^{(2)}=-\frac{\pi\lambda}{2\nu}\left(\frac{2k}{m}\right)^{1/2}
\frac{1}{1-\frac{2\lambda}{\nu}\ln{\frac{m}{k}}}\;,
\end{align*}
where $\boldsymbol{C}$ is the Euler constant.  
An important feature of this mode is the divergence of the 
amplitudes of the fields on the brane in the small mass limit,
\begin{align*}
&\chi_1^{(2)}\big|_{z=0}=-\frac{\mu}{\nu}
\left(\frac{2k}{m}\right)^{1/2}
\frac{1}{1-\frac{2\lambda}{\nu}\ln{\frac{m}{k}}}\;,\\
&\hat{\chi}_2^{(2)}\big|_{z=0}=\frac{\lambda}{\nu}
\left(\frac{2k}{m}\right)^{1/2}
\frac{1}{1-\frac{2\lambda}{\nu}\ln{\frac{m}{k}}}\;.
\end{align*}
Thus, the modes $\chi^{(2)}_m$ give dominant contribution to the
Green's function of the operator (\ref{eitens*}) 
at large distances on the brane. Their collection comprises the
quasilocalized four-dimensional graviton. 

To make the last statement
more precise, let us consider production of gravitational waves by an external 
periodic transverse traceless source on the brane, 
\[
T^{ext}_{ij}(x,z)=\delta(z)\e^{-i\omega t}
\int\frac{d^3{\bf p}}{(2\pi)^3}\e^{i{\bf px}}
d^{(\alpha)}_{ij}({\bf p}) T^{ext}_{\alpha}({\bf p}).
\]
The resulting gravitational field is given by the convolution of the
source with the Green's function 
\be
\label{Greenmix}
G({\bf x-x}';\omega)=-16\pi G_5 \int d\tau~G_{11}(x,x';
z=z'=0)\e^{-i\omega\tau}\;, 
\ee 
where $\tau=t-t'$, and $G_{11}(x,x';z,z')$ is the upper left 
element of the retarded Green's function of
the operator (\ref{eitens*}), 
\be
\label{Green}
G(x,x';z,z')=\int_0^{\infty}dm\sum_{s=1,2}
\begin{pmatrix}
\chi_{1m}^{(s)}(z)\chi_{1m}^{(s)}(z') &
\chi_{1m}^{(s)}(z)\hat{\chi}_{2m}^{(s)}(z')\\
\hat{\chi}_{2m}^{(s)}(z)\chi_{1m}^{(s)}(z')&
\hat{\chi}_{2m}^{(s)}(z)\hat{\chi}_{2m}^{(s)}(z')
\end{pmatrix}
\int\frac{d^4p}{(2\pi)^4}\frac{\e^{-ip(x-x')}}{m^2-p^2-i\epsilon p_0}\;.
\ee
Substituting Eq.~(\ref{Green}) into (\ref{Greenmix}) one obtains,
\be
\label{Greenmix1}
G({\bf x-x}';\omega)=-\frac{4 G_5}{r}\int_0^{\infty}
dm\sum_{s=1,2} \left(\chi_{1m}^{(s)}(0)\right)^2 
\e^{ip_{\omega}r}\;,
\ee
where $r=|{\bf x-x}'|$, $p_{\omega}=\sqrt{\omega^2-m^2}$ when
$m<\omega$ and $p_{\omega}=i\sqrt{m^2-\omega^2}$ when
$m>\omega$. We see that the gravitational field on the brane has the
form of a superposition of massive four-dimensional modes. Only modes
with $m<\omega$ are actually radiated, the other ones exponentially
fall off from the source. Thus, as long as we are interested in the
gravitational waves, we can integrate in (\ref{Greenmix1})
only up to $m=\omega$. This means, in particular, that at $\omega\ll k$
one can neglect the contribution of the modes $\chi^{(1)}_m$ in
Eq.~(\ref{Greenmix1}). 
Let us introduce  
\be
\label{mc}
m_c= k\e^{-\nu/2\lambda}\;,
\ee
and study the regime $m_c\ll\omega\ll k$. The Green's function
(\ref{Greenmix1}) takes the form
\be
\label{Greenmix2}
G({\bf x-x}';\omega)=-\frac{4 G_N}{r}~\e^{i\omega r}~
\cdot\frac{2\lambda}{\nu}\int_0^{\omega} \frac{dm}{m}
\frac{\e^{-i\frac{rm^2}{2\omega}}}
{\left(1-\frac{2\lambda}{\nu}\ln{\frac{m}{k}}\right)^2}\;,
\ee
where $G_N=G_5 k$ is the four-dimensional Newton constant. At 
$r\ll \omega/m_c^2$ the integral in Eq.~(\ref{Greenmix2})
is saturated by $m\sim m_c$ and we obtain the usual four-dimensional
expression for the gravitational wave. We stress that this wave is
a superposition of massive modes with masses of order $m_c$. In the
opposite limit $r\gg \omega/m_c^2$ the integral is damped by the rapidly
oscillating exponent, so one obtains
\[
G({\bf x-x}';\omega)=-\frac{4 G_N}{r}
\frac{\nu}{\lambda\ln{\frac{k^2 r}{2\omega}}}~\e^{i\omega r}\;.
\]
The gravitational wave gradually dissipates into the fifth dimension. 

A few comments are in order. First, the graviton mass scale 
$m_c$ turns out
to be independent of the parameters $\varkappa_1$, $\varkappa_2$ 
in the action
(\ref{action}):
\[
\frac{\nu}{2\lambda}=\frac{1}{16\pi G_5v^2}\;.
\]
This is not very surprising. Indeed, from the very beginning we have
been interested in the modes with masses much smaller than the
energy scale associated with $\varkappa_1$ and 
$\varkappa_2$. Then, the potential term for vector fields 
is effectively frozen at zero and its parameters  
do not affect dynamics. Rather, the condition that the potential
must be zero, generates mixing between gravitons and vectors which
leads to quasilocalization of graviton. The strength of 
mixing at low energies depends only on $v$, and so does the graviton
mass.  

Second, very small values of $m_c$ are generated without
fine-tuning. For example, taking $k, M_5$ as large as the Planck mass and
$v\approx (M_5/5)^{3/2}$ one obtains $m_c$ as small as
$(10^{28}{\rm cm})^{-1}$, that is the inverse of 
the present horizon size of the Universe.

Finally, though the Lorentz symmetry is broken in the background
(\ref{stat*}), the Green's function (\ref{Green}) for the tensor modes
has Lorentz-invariant form.\footnote{This allows to interpret the
additional enhancement of the distance $r\sim \omega/m_c^2$, at which the
dissipation of the gravitational waves sets in, by the factor
$\omega/m_c$ as compared to $m_c^{-1}$, as a relativistic effect, see
Ref.~\cite{Gregory:2000jc}.}  The place where the Lorentz-breaking is
essential is that the very notion of the tensor modes refers to the
preferred reference frame.


\section{Vector modes}

The gauge conditions (\ref{trtr}), (\ref{Agauge}) result in the following
equations for the vector perturbations,
\bseq
\label{trtrvec*}
\begin{align}
\label{trtrvec1}
&p_0\psi_1+{\bf p}^2\psi_2=0\;,\\
\label{trtrvec2}
&p_0\psi_3+{\bf p}^2\psi_4+{\bf p}^2\psi_5=0\;,
\end{align}
\eseq
where we again omitted the polarization index $\alpha$.
From Eqs.~(\ref{hbulkeq}), (\ref{abuleq}) one obtains the 
bulk equations for the vector modes, 
\bseq
\label{vect*}
\begin{align}
\label{vect1}
&\frac{1}{2}\psi_I''-2k^2\psi_I+\frac{p^2}{2u^2}\psi_I=0\;,~~~~I=1,2\;,\\
\label{vect2}
&-\psi_I''+2k\psi_I'-\frac{p^2}{u^2}\psi_I=0\;,~~~~~I=3,4,5~,~~~~~z>0\;,
\end{align}
\eseq
The junction conditions on the brane read:
\bseq
\label{bndvec*}
\begin{align}
\label{bndvec1}
&\psi_1'+2k\psi_1=0\;,\\
\label{bndvec2}
&\psi_2'+2k\psi_2=
\lambda\left(\psi_2+\frac{2\psi_4}{v}+\frac{i\psi_6}{v}\right)\;,\\
\label{bndvec3}
&\psi_3'=\psi_5'=0\;,\\
\label{bndvec4}
&\psi_4'=\varkappa^2v^3
\left(\psi_2+\frac{2\psi_4}{v}+\frac{i\psi_6}{v}\right)\;. 
\end{align}
\eseq
Equations (\ref{bndvec1}), 
(\ref{bndvec2}) together with the gauge fixing condition
(\ref{trtrvec1}) imply
\be
\label{cnstr}
\left.\left(\psi_2+\frac{2\psi_4}{v}\right)\right|_{z=0}+\frac{i\psi_6}{v}=0\;.
\ee 
Thus, the system (\ref{bndvec*}) reduces to a set of homogeneous boundary
conditions for the vector part of the gravitational perturbations and for the free
vectors in the RS background. In other words, mixing between
gravitational and vector fields has no effect in the sector of vector
modes.  The relations (\ref{trtrvec*}),
(\ref{cnstr}) imply three independent vector
modes, parametrized, say, by the functions $\psi_2(z)$, $\psi_3(z)$,
$\psi_4(z)$.

Let us show the absence of physical zero modes in the vector
sector. Assume that there is a normalizable mode with $p^2=0$.
Then, from Eqs.~(\ref{vect*}) we find that 
$\psi_2=C_2\e^{-2k|z|}$, and that $\psi_3$, $\psi_4$
do not depend on $z$. The latter fact means that 
the corresponding vector $a_{\mu}^a$ is independent of $z$, see
Eq.~(\ref{decompaai}). Consequently the 
vector field
strength $F_{\mu\nu}^a$ 
is also independent of $z$, and, for the mode
to be normalizable, $F_{\mu\nu}^a$ must be zero. This implies that the
vector field $a_{\mu}^a$ of the zero mode is pure gauge in the bulk.
From Eqs.~(\ref{decomph0i}), 
(\ref{decomphij}) we find that the metric of the zero
mode is also pure gauge,
\[
h_{\mu\nu}(x,z)=(\xi_{\mu,\nu}+\xi_{\nu,\mu})\e^{-2k|z|}~,~~~~~
\xi_{\mu}=-iC_{2\alpha}e_{\mu}^{(\alpha)}\e^{-ipx}\;.
\]
Thus, in the bulk, this mode can be removed by residual gauge transformation
left after fixing the transverse-traceless gauge (\ref{trtr}),
(\ref{Agauge}). The only degree of freedom we are left with is the vector
component $\psi_6$ of the field $\beta^a$, see Eq.~(\ref{decompbetaa}).
This field resides on the
brane and 
describes explicit breaking of the $[U(1)]^3$ gauge
symmetry by the brane action, cf. Eq.~(\ref{aGN}).
But $\psi_6$ vanishes due to
Eq.~(\ref{cnstr}). Hence, all components of the mode vanish: 
there is no normalizable zero mode at all.   
One concludes that there is only the continuum
spectrum in the vector sector with $p^2\geq 0$. 
The corresponding modes are completely
delocalized and have the same form as in the pure RS case.

It is worth mentioning that at $p^2=0$ the mode $\psi_6$ 
vanishes due to field
equations and boundary conditions. This could be dangerous, if this
mode
became propagating in a non-trivial background. As we show in
Appendix B, this is not the case, which implies also that 
$\psi_6$ remains non-dynamical to
higher orders in classical perturbation theory. 


\section{Scalar sector}
We now turn to the scalar sector. Having in mind the 
issue of vDVZ discontinuity, we will consider 
the gravitational field produced on the
brane by an external energy-momentum 
tensor $T^{ext}_{\mu\nu}$ localized on the brane. We do not restrict
our analysis to static sources and consider external energy-momentum
tensor depending both on time and space.
Passing to the 4-dimensional Fourier representation
we decompose it as follows,
\[
T^{ext}_{00}=t_1\;,~~~~
T^{ext}_{0i}=p_it_2+\ldots\;,~~~~
T^{ext}_{ij}=\left(\delta_{ij}-\frac{p_ip_j}{{\bf p}^2}\right)t_3+
p_ip_jt_4+\ldots\;,
\]
where we wrote down explicitly only the scalar components, which are
of interest to us. 
The energy-momentum conservation implies 
\[
p_0t_1+{\bf p}^2t_2=0\;,~~~~
p_0t_2+{\bf p}^2t_4=0\;.
\]
Below we will consider two gauge-invariant scalar potentials,
$\bar{\phi}_3$ and
$\bar{\Phi}\equiv\bar{\phi}_1+2p_0\bar{\phi}_2+p_0^2\bar{\phi}_4$, of
the metric produced by the external source on the brane.  The
quantities with the bar are defined via decomposition of the metric on
the brane in accordance with (\ref{decomph00})--(\ref{decomphij}).
The potentials $\bar{\phi}_3$, $\bar{\Phi}$ are related to the
standard Bardeen variables
$\Phi$, $\Psi$ (in
the notations of Ref.~\cite{Mukhanov:1990me}) as follows,
$\bar{\phi}_3=2\Psi$, $\bar{\Phi}=2\Phi+\frac{2p_0^2}{{\bf p}^2}\Psi$.
Using the
relation (\ref{hGN}) between the metric induced on the brane and the
bulk metric we obtain,
\begin{gather}
\label{phibarphi1}
\bar{\phi}_3=\phi_3\big|_{z=0}+2k\varepsilon\;,\\
\label{phibarphi2}
\bar{\Phi}=(\phi_1+2p_0\phi_2+p_0^2\phi_4)\Big|_{z=0}
+\frac{2kp^2}{{\bf p}^2}\varepsilon\;.
\end{gather}
Our purpose is to compute the potentials $\bar{\phi}_3$ and $\bar{\Phi}$
and to compare them to the expressions obtained in the usual
four-dimensional Einstein gravity.

In the bulk, the fields obey the following equations, see 
Eqs.~(\ref{hbulkeq}), (\ref{abuleq}),
\bseq
\label{scal*}
\begin{align}
\label{scal1}
&\frac{1}{2}\phi_I''-2k^2\phi_I+\frac{p^2}{2u^2}\phi_I=0\;,~~~~I=1,2,3,4\;,\\
\label{scal2}
&-\phi_I''+2k\phi_I'-\frac{p^2}{u^2}\phi_I=0\;,~~~~~I=5,6,7~,~~~~~z>0\;,
\end{align}
\eseq
The gauge conditions 
(\ref{trtr}), (\ref{Agauge}) take the form,
\bseq
\label{scalconst*}
\begin{gather}
\label{scalconst1}
\phi_1-2\phi_3-{\bf p}^2\phi_4=0\;,\\
\label{scalconst2}
p_0\phi_1+{\bf p}^2\phi_2=0\;,\\
\label{scalconst3}
p_0\phi_2+{\bf p}^2\phi_4=0\;,\\
\label{scalconst4}
p_0\phi_5+{\bf p}^2\phi_7=0\;.
\end{gather} 
\eseq 
As described in Appendix A, the junction conditions 
in the presence of the external
energy-momentum tensor on the brane can be
cast into the following form, 
\bseq
\label{scfin*}
\begin{align}
&\phi_1'+2k\phi_1=2{\bf p}^2\varepsilon+8\pi G_5t_1\;,
\label{scfin1}\\
&\phi_2'+2k\phi_2=-2p_0\varepsilon+8\pi G_5t_2\;,
\label{scfin2}\\
&\phi_3'+2k\phi_3=-p^2\varepsilon-4\pi G_5\frac{p^2}{{\bf p}^2}t_1\;,
\label{scfin3}\\
&\phi_4'+2k\phi_4=\frac{2p_0^2}{{\bf p}^2}\varepsilon+8\pi G_5t_4\;,
\label{scfin4}\\
&\phi_5'=0\;,
\label{scfin5}\\
&\phi_6'=-\frac{3\nu_1}{2\lambda_1}vp^2\varepsilon-
2\pi G_5\frac{\nu_1}{\lambda_1}v(-t_1+2t_3+{\bf p}^2 t_4)\;,
\label{scfin6}\\
&\phi_7'=0\;,
\label{scfin7}\\
&-p^2\varepsilon=
\lambda_1\rho\left(\phi_3+\frac{2}{v}\phi_6+2k\varepsilon\right)
+\frac{4\pi G_5}{3}(-t_1+2t_3+{\bf p}^2t_4)\;,
\label{scfin8}
\end{align}
\eseq 
where 
\[
\nu_1=2\varkappa_1^2v^2\;,~~~~~~ 
\rho=\frac{\lambda_1+\lambda_2}{3\lambda_1+2\lambda_2}\;.
\]
These equations should be supplemented by Eq.~(\ref{sccomb1}) in
Appendix A, which
determines the longitudinal component $\phi_8$ in terms of the other
fields. Note that Eqs.~(\ref{scfin3}), (\ref{scfin6}), (\ref{scfin8})
form a closed system. Once the solution of this subsystem is
found, the other equations are solved in a straightforward manner.  

Before proceeding with the calculation of the gravitational field
produced by the source, let us make sure that the scalar sector does
not contain instabilities. So, one temporarily sets the external source
equal to zero in Eqs.~(\ref{scfin*}).  First, we show that there is no
tachyonic mode, which would correspond to negative or complex $p^2$.  
For such a mode,
let
us introduce $w$, such that $k^2w^2=-p^2$ and $\Re{w}>0$. Then, for a
normalizable mode, one would have 
\[
\phi_3=U_3 K_2(w\e^{k|z|})~,~~~~\phi_6=U_6 w\e^{k|z|} K_1(w\e^{k|z|})\;,
\] 
where $K_1$, $K_2$ are modified Bessel functions.
From Eqs.~(\ref{scfin3}), (\ref{scfin6}) we obtain
\[
U_3=-\frac{kw\varepsilon}{K_1(w)}~,~~~~
U_6=-\frac{3\nu_1 vk\varepsilon}{2\lambda_1 K_0(w)}\;.
\]
Inserting these expressions into Eq.~(\ref{scfin8}) one obtains
the
following relation,
\be
\label{tachx1}
w+\frac{\lambda_1\rho}{k}\left[\frac{K_0(w)}{K_1(w)}+
\frac{3\nu_1}{\lambda_1}\frac{K_1(w)}{K_0(w)}\right]=0\;.
\ee
Let us show that the real part of the expression in square brackets is
positive in the right half-plane $\Re w>0$. Indeed, on the boundary of
this half-plane
one has,
\begin{align*}
&\Re\left[\frac{K_1(w)}{K_0(w)}\right]\bigg|_{w=iy} =\frac{2}{\pi
|y|(J_0^2(y)+N_0^2(y))}>0\;,\\
&\Re\left[\frac{K_1(w)}{K_0(w)}\right]\bigg|_{|w|\to\infty,-\pi/2\leq
\mathrm{Arg}\, w \leq \pi/2}=1>0\;.
\end{align*} 
Hence, $\Re\left[\frac{K_1(w)}{K_0(w)}\right]$, being a harmonic
function, is positive everywhere inside the half-plane $\Re
w>0$. This ensures the positivity of the real part of the inverse
function, $\Re\left[\frac{K_0(w)}{K_1(w)}\right]>0$, and thus of the whole
expression in square brackets in Eq.~(\ref{tachx1}). One concludes
that Eq.~(\ref{tachx1}) has no solutions in the right half-plane,
implying the absence of tachyonic modes in the scalar sector.

Second, let us demonstrate that the model is also free of ghosts. The ghost
mode, if any, must be localized on the brane. Indeed, the
normalization of the modes of continuum spectrum is determined
entirely by the bulk action, which by itself is free of ghosts.  It is
straightforward to see that the modes corresponding to strictly
positive $p^2$ belong to continuum part of the spectrum. The only
dangerous eigenvalue is $p^2=0$. 
But  
the corresponding mode is unphysical. Indeed, 
for this mode
 $\phi_2=U_2\e^{-2k|z|}$ and from
Eq~(\ref{scfin2}) one obtains $\varepsilon=0$.
Substituting ${\bf p}^2=p_0^2$ into the gauge fixing conditions
(\ref{scalconst2}), (\ref{scalconst3}) one finds
$\phi_1=-p_0U_2 \e^{-2k|z|}$, $\phi_4=-\frac{U_2}{p_0} \e^{-2k|z|}$.
Then, Eq.~(\ref{scalconst1}) yields
$\phi_3=0$. 
It follows from Eq.~(\ref{scal2}) that the 
scalar component $\phi_7$ of the vector fields does not depend on
$z$. From Eq.~(\ref{scalconst4}) we find $\phi_5=-p_0 \phi_7$.   
Finally, from Eq.~(\ref{scfin8}) we obtain $\phi_6=0$.
The surviving mode is  
pure gauge in the bulk, see Eqs.~(\ref{decomp*}), 
\begin{align*}
&h_{\mu\nu}=(\xi_{\mu,\nu}+\xi_{\nu,\mu})\e^{-2k|z|}~,~~~~
\xi_{\mu}=\frac{p_{\mu}}{2ip_0}U_2\e^{-ipx}\;,\\
&a^a_{\mu}=\d_{\mu}\left(-ip_a \phi_7\e^{-ipx}\right)\;.
\end{align*}
The bulk fields can be removed by a residual gauge transformation,
leaving the scalar component $\phi_8$ (see Eq.~(\ref{decompbetaa}))  
of the fields $\beta^a$
on the brane. But the latter
is zero according to the field equation 
(\ref{sccomb1}). Thus, all the
components of the zero mode 
vanish. There is only
continuum spectrum of modes with $p^2>0$ in the scalar sector. 

We show in Appendix B that the mode $\phi_8$ remains
non-dynamical in a non-trivial background as well. Hence, no new
propagating degrees of freedom appear both in non-trivial backgrounds
and in higher orders of classical perturbation theory.

Having established the absence of instabilities, 
we now return to the field produced by the source. Our strategy is to
express the functions $\phi_3(z)$, $\phi_6(z)$ in terms of the external
source and the brane bending $\varepsilon$ using the bulk equations with
the boundary conditions (\ref{scfin3}), (\ref{scfin6}), and to insert the
result into Eq.~(\ref{scfin8}). This is   
most easily done by combining the bulk equations with the boundary
conditions into the Schr\"odinger type operators and introducing the
Green's functions of these operators. The latter satisfy the
following equations,
\begin{align*}
-u^2G_3''(z;p)+4k^2 u^2G_3(z;p)-4k\delta(z)G_3(z;p)
-p^2G_3(z;p)=\delta(z)\;,\\
-u^2G_6''(z;p)+2ku^2G_6'(z;p)-p^2G_6(z;p)=\delta(z)\;.
\end{align*}
Imposing the radiation (outgoing wave) boundary conditions at $z\to
\pm\infty$ one obtains, 
\begin{gather}
\label{Green3}
G_3(z;p)=-\frac{H_2^{(1)}(p\e^{k|z|}/k)}{2pH_1^{(1)}(p/k)}\;,\\
\label{Green6}
G_6(z;p)=-\frac{\e^{k|z|} H_1^{(1)}(p\e^{k|z|}/k)}{2pH_0^{(1)}(p/k)}\;,
\end{gather}
where $p=\sqrt{p^2}$, $\Re p\geq 0$. Making use of these Green's
functions we find,
\begin{align} 
\label{phi30}
&\phi_3(0)=
-\frac{pH_2^{(1)}(p/k)}{H_1^{(1)}(p/k)}
\left(\varepsilon+4\pi G_5\frac{t_1}{{\bf p}^2}\right)\;,\\
\label{phi60}
&\phi_6(0)=-\frac{H_1^{(1)}(p/k)}{pH_0^{(1)}(p/k)}
\left(\frac{3\nu_1}{2\lambda_1}vp^2\varepsilon+
2\pi G_5\frac{\nu_1}{\lambda_1}v(-t_1+2t_3+{\bf p}^2t_4)\right).
\end{align}
Substitution of these expressions into Eq.~(\ref{scfin8}) yields,
\be
\label{epsil}
\begin{split}
&\varepsilon\left[p^2+\lambda_1\rho\frac{pH_0^{(1)}(p/k)}{H_1^{(1)}(p/k)}
-3\nu_1\rho \frac{pH_1^{(1)}(p/k)}{H_0^{(1)}(p/k)}\right]\\
&-\frac{4\pi G_5}{3}T
\left[1-3\nu_1\rho\frac{H_1^{(1)}(p/k)}{pH_0^{(1)}(p/k)}\right]
-4\pi G_5\lambda_1\rho\frac{pH_2^{(1)}(p/k)}{H_1^{(1)}(p/k)}
\frac{t_1}{{\bf p}^2}=0
\end{split}
\ee
where we have used the relation
\[
2k-\frac{pH_2^{(1)}(p/k)}{H_1^{(1)}(p/k)}=
\frac{pH_0^{(1)}(p/k)}{H_1^{(1)}(p/k)}
\]
and introduced the notation $T=t_1-2t_3-{\bf p}^2t_4$ for the trace of
the energy-momentum tensor. 

Further analysis depends on the value of the parameter
$\varkappa_1^2$. Let us first consider the case $\varkappa_1^2=0$, which in
Eq.~(\ref{action}) corresponds to the absence of the first term in the  
potential for vector fields.\footnote{We do not
know whether vanishing of the parameter $\varkappa_1^2$ can be ensured
by any symmetry requirement.} 
In this case $\lambda_1=\nu_1=0$, and 
Eq.~(\ref{epsil}) reduces to, 
\be
\label{epsilRS}
\varepsilon=\frac{4\pi G_5}{3p^2} T\;.  
\ee 
The last expression coincides with the result obtained in the pure RS case,
cf. \cite{Garriga:1999yh}. Substitution of (\ref{epsilRS}) into
Eqs.~\eqref{phi30}, (\ref{phi60}) leads\footnote{Note that though
$\lambda_1=\nu_1=0$, the ratio $\frac{\nu_1}{\lambda_1}=\frac{1}{8\pi
G_5v^2}$ in (\ref{phi60}) is finite.}  to vanishing of the scalar part of
the vector fields, $\phi_6=0$, while the scalar components of the
metric
are the same as in the pure RS case. Hence, there are only
short-distance corrections to the scalar part of the metric in the
case $\varkappa_1^2=0$. At distances $r\gg 1/k$, i.e. at small momenta
$p\ll k$, one arrives at the same expressions as in the four-dimensional
Einstein gravity, 
\be
\label{potfour}
\bar{\phi}_3=-\frac{8\pi G_N}{{\bf p}^2}t_1~,~~~~~~
\bar{\Phi}=\frac{8\pi G_Np^2}{({\bf p}^2)^2}t_1
-\frac{16\pi G_N}{{\bf p}^2}t_3\;,
\ee
where $G_N=G_5k$ is the four-dimensional Newton constant. 
Evidently, the gravitational field is free from the vDVZ
discontinuity. In particular, the Newton law --- the 
field of a static source --- does not get
modified at large distances in the case $\varkappa_1^2=0$.

Another possibility is that the parameter $\varkappa_1^2$ is large. In
this case we are interested in the behavior of the fields at distances
$r\gg 1/\nu_1$, i.e., at small values of the
momentum, $p\ll k, \nu_1$. We also assume as in Sec.~3 the hierarchy
$\lambda_1\ll\nu_1$. The dominant contributions to the expressions in
the square brackets in Eq.~(\ref{epsil}) come from the terms
proportional to $H^{(1)}_1(p/k)/H^{(1)}_0(p/k)$. It is worth noting 
that these terms
come from the scalar component $\phi_6$ of the vector fields in
Eq.~(\ref{scfin8}). 
Keeping only the leading contributions at low momenta, one obtains
from
Eq.~(\ref{epsil}),
\be
\label{epsil5}
\varepsilon=\frac{4\pi G_5}{3p^2}T+\frac{8\pi G_5}{3{\bf p}^2}t_1
\frac{\lambda_1}{\nu_1}\ln{\frac{p}{k}}\;.
\ee
This expression for the brane bending differs from the formula
(\ref{epsilRS}) in the pure RS case by the last term, which is small
at moderate momenta. However, it becomes important 
and is
logarithmically large relative to the first term in the far infrared.
Making use of the expression (\ref{epsil5}) 
for the brane bending $\varepsilon$ and the
Green's function (\ref{Green3}) one determines the rest of the
gravitational modes, $\phi_1$, $\phi_2$, $\phi_4$. Finally, these
expressions, being inserted into
Eqs.~(\ref{phibarphi1}), (\ref{phibarphi2}), yield the scalar
components of the metric induced on the brane,
\begin{align}
\label{phibar3}
&\bar{\phi}_3=-8\pi G_N\frac{t_1}{{\bf p}^2}\;,\\
\label{barPhi2}
&\bar{\Phi}=
\frac{8\pi G_N p^2}{({\bf p}^2)^2}t_1-\frac{16\pi G_N}{{\bf p}^2}t_3
+\frac{16\pi G_N p^2}{({\bf p}^2)^2}t_1\frac{\lambda_1}{\nu_1}
\ln{\frac{p}{k}}\;.
\end{align}
The
expression (\ref{phibar3}) coincides with the result of the usual
four-dimensional Einstein gravity. Thus, the potential $\phi_3$ does
not get modified at large distances at all. As to the second
potential, the first two terms in Eq.~(\ref{barPhi2}) also coincide
with the result of conventional gravity. The last
term provides the long-distance correction, which becomes important at
the distance 
\[
r_c=\frac{1}{k}\e^{\nu_1/2\lambda_1}=\frac{1}{k}\e^{1/16\pi G_5v^2}\;. 
\]
This distance coincides with $1/m_c$, where $m_c$ is the
graviton mass scale (\ref{mc}).

To make the physical consequences of the formula (\ref{barPhi2}) more
clear, let us calculate the gravitational field produced by a
point-like static source. We take $T_{00}=M\delta({\bf x})$,
$T_{0i}=T_{ij}=0$. Setting $t_1=M$, $t_3=0$, $p^2=-{\bf p}^2$ in
Eq.~(\ref{barPhi2}) and performing Fourier transform, one obtains, \be
\bar{\Phi}(r)=-\frac{2G_N M}{r}
\left(1-\frac{2\lambda_1}{\nu_1}\ln{kr}\right)\;.
\label{Newton}
\ee
while for the other gauge invariant potential we have,
\be
\bar{\phi}_3(r)=-\frac{2G_N M}{r}\;.
\label{Newton3}
\ee 
These expressions are valid up to corrections at small distances
$r\sim 1/k, 1/\nu_1$. 
The potential \eqref{Newton} is the analog of the Newton potential
in our model, it is responsible for gravitational
interaction between nonrelativistic massive objects. 
Remarkably, a contribution to
$\bar{\Phi}(r)$ appears, which grows logarithmically with the distance
as 
compared to the standard expression $(-2G_N M/r)$. Even more strikingly,
the sign of this contribution is opposite to that of 
the standard expression. As
a result, the expression \eqref{Newton} describes gradual
weakening of gravitational attraction with the distance;  
at large
distances, $r>1/m_c$, attraction gets replaced by repulsion.
Thus, our model provides an example
of a ghost-free theory with antigravity at ultra-large distances. 

At first sight it seems surprising that in the model with
quasilocalized gravitons we obtain a contribution to the Newton
potential which grows at large distances, as compared to the standard
four-dimensional expression. Following the conventional line of
reasoning one could infer that, as gravitons dissipate at large
distances into the fifth dimension, the gravitational interaction
should become weaker at large distances, as compared 
to the four-dimensional
case. However, this line of reasoning is incorrect. In theories with
Lorentz symmetry breaking, the potential produced by an external source
is not directly related to the spectrum of propagating degrees of
freedom. This point is illustrated by four-dimensional Lorentz
violating massive electrodynamics, considered in
Refs.~\cite{Gabadadze:2004iv}. In that case the electric potential of
static sources falls off as $1/r$ in spite of the fact that all 
propagating modes are massive.

The origin of the logarithmically enhanced antigravity in our model can be
understood as follows. A point mass gives rise to perturbations of the
vector fields which interact with matter via mixing with the
metric. The gravitational field is produced by the total
energy-momentum tensor composed of $T^{ext}_{\mu\nu}$ and the
energy-momentum tensor $T^V_{\mu\nu}$ of the vectors, given by
Eq.~(\ref{Tmunu}). The latter tensor falls off slowly from the
localized external source, and dominates at large distances. Our
results indicate that, insofar as the Newton potential is concerned,
the vector fields mimic the effect of negative energy. 
This is possible because
$T^V_{\mu\nu}$ violates the weak energy condition. 

The gravitational potentials (\ref{phibar3}), (\ref{barPhi2}) are free
from the vDVZ discontinuity. 
Indeed, in the limit of vanishing graviton mass scale $m_c$, which
corresponds to $\lambda_1/\nu_1\to 0$, one recovers
the usual four dimensional expressions
(\ref{potfour}). However, from the phenomenological point of view the
difference between the two gravitational potentials (\ref{Newton}) and
(\ref{Newton3}) results in a severe phenomenological constraint on the
value of the parameter $\lambda_1/\nu_1$. Measurements of the light
deflection by the gravitational field of the Sun \cite{lightdefl}
require $\lambda_1/\nu_1 \lsim 10^{-5}$. This value is not unnaturally
small: it corresponds to $v\sim (0.027M_5)^{3/2}$.  However, it pushes
the range $1/m_c$, where the antigravity sets in, far beyond the
present horizon size of the Universe. The mass scale $m_c$ in this
case is very small and it is not clear whether it can significantly
affect the physics within the present horizon.  We discuss possible
ways of avoiding this phenomenological constraint in the concluding
section.


\section{Discussion}
In this paper we presented a Lorentz-violating 
brane-world model, where gravitons are not
completely localized, but are rather quasilocalized on the brane. In
other words, the four-dimensional graviton is a collection of KK modes
from the continuum spectrum. The characteristic mass $m_c$ of these
modes is exponentially small when expressed in terms of the 
parameters of the Lagrangian. 
We demonstrated that the model is
free from tachyonic and ghost-like instabilities. We calculated the
metrics produced by an external energy-momentum tensor on the brane,
and found that the model does not suffer from the vDVZ
discontinuity.

The key observation behind the model 
is that while the gravitational perturbations in the Randall--Sundrum setup
contain a zero mode localized on the
brane, the bulk vector fields are completely delocalized.  
Thus, mixing between the vectors and the would-be
zero gravitational mode forces the latter to dissipate into the fifth
dimension. 
This mixing between tensors and vectors becomes possible when the 
Lorentz symmetry is spontaneously broken by 
non-zero VEVs of the vector fields. These considerations are rather
generic. We believe them to be applicable to a broad class of
generalizations of the Randal--Sundrum model with 
spontaneous Lorentz symmetry breaking by
bulk fields.
In particular, the vector fields can be replaced by form-fields 
of higher degrees, which are also completely delocalized in the
Randall--Sundrum background.

For general form of the potential term for the vector fields 
on the brane the model
exhibits antigravity at ultra-large
distances. Namely, the structure of the Newton potential is such
that gravitational attraction between nonrelativistic massive objects
gradually weakens as the distance increases, and gets replaced by
repulsion at $r>1/m_c$. While this property is theoretically
appealing, it puts severe phenomenological constraints on the
model. The reason is that the second Bardeen potential,
characterizing spatial part of the metric of a static source, does not get
modified at large distances and thus differs from the Newton
potential. Observations of light deflection by the Sun constrain the
relative difference between the two gravitational potentials at the
level of $10^{-4}$. This translates into the constraint 
$m_c\lsim M_{Pl}\exp{(-10^5)}$. This value is negligible compared 
to
the present Hubble parameter of the Universe. It is doubtful
whether the model with so tiny graviton mass scale can lead to any
interesting phenomenology, in particular, to accelerating expansion of
the Universe
at the present epoch.

One possibility to avoid this phenomenological problem
is to fine-tune the parameter
$\varkappa_1$ of the vector potential in the action (\ref{action}) to
zero. Then, both Bardeen potentials have at large
distances the same form as in the four-dimensional 
linearized Einstein gravity, while
graviton is still quasilocalized and gravity waves dissipate into
extra dimensions. The graviton mass scale $m_c$ in this case can be
comparable to the Hubble parameter or even larger. 
A drawback of this approach is fine-tuning of $\varkappa_1$. We are
not aware whether
vanishing of this parameter can be imposed by any symmetry.

Another way out is a generalization of the model considered in this
paper. Let us sketch a particular example.
One introduces a scalar field with dilaton-like
coupling to the vector fields on the brane. Namely, one replaces 
the combination 
$\bar{g}^{\mu\nu}A^a_{\mu}A^b_{\nu}$ in the brane action in
Eq.~(\ref{action}) by
$\bar{g}^{\mu\nu}\e^{\alpha\varphi}A^a_{\mu}A^b_{\nu}$. 
The modified model allows for the Lorentz-violating background
(\ref{stat*}) with $\varphi=0$. 
The dilaton
does not affect tensor and vector sectors of the linearized
perturbation above this background, so the modified 
model also incorporates
quasilocalized gravitons. On the other hand, relative difference
between Bardeen potentials depends on the dilaton coupling, and is
essentially proportional to $1/\alpha^2$ when $\alpha$ is large. Thus, the
constraint from light deflection is satisfied once $\alpha\gsim 100$. 
We will report more on phenomenology of our model and its
generalizations elsewhere. 

Two other important issues which 
we leave for future investigations are the 
quantum structure of the theory, 
and cosmology in the model (\ref{action}) and its
generalizations. 
Of special interest is at what scale the strong coupling at
the quantum level sets in, and what kind of late-time 
cosmological evolution is obtained in these models, in
particular,  
whether they can account for the 
cosmic acceleration at the present epoch.

\paragraph*{Acknowledgements}
We are indebted to S.~Dubovsky, 
M.~Libanov and V.~Rubakov for many fruitful
discussions and helpful suggestions. We are grateful to S.~Demidov,
D.~Krotov, D.~Levkov, Kh.~Nirov, E.~Nugaev for their encouraging
interest. 
This work has been supported in part by the grant of the President of
the Russian Federation NS-2184.2003.2, and the Russian Foundation for
Basic Research grant 05-02-17363. The work of D.G. has been supported
in part by the RFBR grant 04-02-17448, 
INTAS grant 03-51-5112, the grant of
the Russian Science Support Foundation and by the 
fellowship of the "Dynasty" foundation (awarded by the Scientific
board of ICFPM).

  
\appendix
\section{Junction conditions in the scalar sector}

In this Appendix we
consider the junction conditions for the scalar modes in the presence
of an external energy-momentum tensor on the brane. The external source 
should be added to
the r.h.s. of Eqs.~(\ref{hbound*}) according to
Eq.~(\ref{boundmunu}). As a result one obtains 
the following junction
conditions on the brane,
\bseq
\label{bboundscal*}
\begin{align}
\notag
\phi_1'+2k\phi_1-2p_0^2\varepsilon=&
\frac{\lambda_1}{3}\bigg(2\phi_3+{\bf p}^2\phi_4+\frac{4}{v}\phi_6
+\frac{2{\bf p}^2}{v}\phi_7-\frac{{\bf p}^2}{k}\varepsilon+6k\varepsilon
+\frac{2i{\bf p}^2}{v}\phi_8 \bigg)\\
&+8\pi G_5\bigg(\frac{2}{3}t_1+\frac{2}{3}t_3+\frac{{\bf
p}^2}{3}t_4\bigg)\;,
\label{bboundscal1}\\
\phi_2'+2k\phi_2+2p_0\varepsilon=&8\pi G_5 t_2\;,
\label{bboundscal2}
\end{align}
\begin{align}
\label{bboundscal3}
\phi_3'+2k\phi_3=\lambda\bigg(\frac{1}{3}\phi_3&-
\frac{{\bf p}^2}{3}\phi_4+\frac{2}{3v}\phi_6
-\frac{2{\bf p}^2}{3v}\phi_7+\frac{{\bf p}^2}{3k}\varepsilon
-\frac{2i{\bf p}^2}{3v}\phi_8\bigg)\;,\\
\notag
\phi_4'+2k\phi_4-2\varepsilon=\lambda&\bigg(
-\frac{2}{3{\bf p}^2}\phi_3+\frac{2}{3}\phi_4
-\frac{4}{3{\bf p}^2v}\phi_6+\frac{4}{3v}\phi_7-
\frac{2}{3k}\varepsilon+\frac{4i}{3v}\phi_8\bigg)\\
&+8\pi G_5\bigg(\frac{1}{3{\bf p}^2}t_1-\frac{2}{3{\bf p}^2}t_3
+\frac{2}{3}t_4\bigg)\;,
\label{bboundscal4}
\end{align}
\begin{align}
\phi_5'=&0\;,
\label{bboundscal5}\\
\phi_6'=&\varkappa_1^2v^3\bigg(\phi_3+\frac{2}{v}\phi_6+2k\varepsilon\bigg)
+\varkappa_2^2v^3\bigg(\frac{1}{3}\phi_3-
\frac{{\bf p}^2}{3}\phi_4+\frac{2}{3v}\phi_6
-\frac{2{\bf p}^2}{3v}\phi_7+\frac{{\bf p}^2}{3k}\varepsilon
-\frac{2i{\bf p}^2}{3v}\phi_8\bigg)\;,
\label{bboundscal6}\\
\notag
\phi_7'=&\varkappa_1^2v^3\bigg(\phi_4+\frac{2}{v}\phi_7-
\frac{1}{k}\varepsilon+\frac{2k}{{\bf p}^2}\varepsilon
+\frac{2i}{v}\phi_8\bigg)\\
&+\varkappa_2^2v^3\bigg(
-\frac{2}{3{\bf p}^2}\phi_3+\frac{2}{3}\phi_4
-\frac{4}{3{\bf p}^2v}\phi_6+\frac{4}{3v}\phi_7-
\frac{2}{3k}\varepsilon+\frac{4i}{3v}\phi_8\bigg)
\label{bboundscal7}\;.
\end{align}
\eseq
Combining these equations with the gauge fixing conditions
(\ref{scalconst*}), we obtain
\bseq
\label{sccomb*}
\begin{align}
\notag
\lambda_1\bigg\{\bigg(-{\bf p}^2&\phi_4-\frac{2{\bf p}^2}{v}\phi_7
\bigg)\bigg|_{z=0}
+\frac{{\bf p}^2}{k}\varepsilon-2k\varepsilon
-\frac{2i{\bf p}^2}{v}\phi_8\bigg\}\\
&+\lambda_2\bigg\{\bigg(\frac{2}{3}\phi_3-\frac{2{\bf p}^2}{3}\phi_4
+\frac{4}{3v}\phi_6-\frac{4{\bf p}^2}{3v}\phi_7\bigg)\bigg|_{z=0}
+\frac{2{\bf p}^2}{3k}\varepsilon-\frac{4i{\bf p}^2}{3v}\phi_8\bigg\}=0\;,
\label{sccomb1}\\
-2p^2\varepsilon=&\frac{\lambda_1}{3}\left\{\bigg(2\phi_3
+{\bf p}^2\phi_4+\frac{4}{v}\phi_6+\frac{2{\bf p}^2}{v}\phi_7\bigg)
\bigg|_{z=0}
-\frac{{\bf p}^2}{k}\varepsilon+6k\varepsilon
+\frac{2i{\bf p}^2}{v}\phi_8\right\}\notag\\
&+8\pi G_5\left(-\frac{1}{3}t_1+\frac{2}{3}t_3
+\frac{{\bf p}^2}{3}t_4\right)\;. 
\label{sccomb2}
\end{align}
\eseq
The first of these equations plays the role similar to that of
Eq.~(\ref{cnstr}) in the vector sector: it determines the field
$\phi_8$ in terms of the other fields. Substitution of
Eqs.~(\ref{sccomb*}) back into Eqs.~(\ref{bboundscal*}) yields the
system (\ref{scfin*}) considered in the main text.


\section{Absence of ghosts above a non-trivial background}

We saw in the main text that the modes 
$\phi_8$ and $\psi_{6\alpha}$, accounting for the explicit $[U(1)]^3$
gauge symmetry breaking on the brane, vanish the the linear
order in perturbation theory above the background (\ref{stat*}).
These vanishing modes are generically dangerous.
In higher orders of classical perturbation theory (i.e. above
non-trivial backgrounds) kinetic terms for these modes may arise,
rendering them propagating. Depending on the sign of kinetic term some
of these modes may become ghosts.
To study this issue
we investigate the ultraviolet behavior of the 
perturbations of the vector fields above a non-trivial vector
background close to the static background $A_i^a=v\delta^a_i$. For the
sake of simplicity we neglect gravity perturbations. This can be done
consistently by taking the limit
$G_5\to 0$; we believe that switching on gravity 
does not spoil our results. The non-trivial background is chosen
to be locally space-like $A_\mu^a=(0,{\bf A}^a)$.  
In this Appendix we show that the equations of motion for the linear
perturbations above the non-trivial vector background imply vanishing
of the dangerous modes, in complete analogy to the situation above the
static background (\ref{stat*}). This demonstrates the ultraviolet
stability of the model (\ref{action})  
to higher orders in classical perturbation theory. 

Let us consider 
the brane part of the quadratic action for the fluctuations $a_\mu^a$
of the vector fields above the non-trivial vector background,
\begin{align*}
S^{(2)}=&\int d^4x\Bigl[ 
-2\l\varkappa_1^2+\varkappa_2^2\r \l 
A^{\mu~b}A_\mu^c a^{\nu~b}a_\nu^c+A^{\mu~b}a_\mu^c a^{\nu~b}A_\nu^c 
+A^{\mu~b}a_\mu^c a^{\nu~c}A_\nu^b\r\\ 
&-2\varkappa_1^2v^2\cdot a^{\nu~b}a_\nu^b 
-\frac{2\varkappa^2_2}{3}\l
A^{\mu~b}A_\mu^b a^{\nu~c}a_\nu^c+
2A^{\mu~b}a_\mu^bA^{\mu~c}a_\mu^c
\r\Bigr]\;,
\end{align*}
where summation over internal indices $b,c=1,\dots,3$ is assumed, and
the metric on the brane is taken to be flat. The bulk part of the
action is irrelevant for us as we are interested in longitudinal modes 
$a_\mu^a=\d_\mu\beta^a$ which are pure gauge in the bulk. 

Let us
consider equations for the spatial components of the fluctuations,
$a_i^a=\d_i\beta^a$. It is convenient to represent them in the
matrix form introducing $3\times3$ matrices $\hat{a} = a_i^a$ and
$\hat A =A_i^a$. By a suitable transformation, the background matrix
$\hat A$ can be made symmetric, $\hat A= \hat A^T$.
Then the equation of motion for
the fluctuations reads,
\begin{equation}
4\l\varkappa_1^2+\varkappa_2^2\r \l
\hat A^2 \hat a + \hat A \hat a^T \hat A + \hat a \hat A^2
\r
+4\varkappa_1^2v^2\cdot\hat a
\label{matrix-equations}
+
\frac{4\varkappa^2_2}{3}
\l
2{\rm Tr}[\hat a\hat A]\cdot \hat A+ 
{\rm Tr}[\hat A^2]\cdot \hat a
\r=0\;.
\end{equation}
The
symmetric part of the fluctuations, $s_i^a=\d_i\beta^a+\d_a\beta^i$, 
obeys the homogeneous equation,
\[
{\cal A}[\hat s]=0\;,
\]
where the linear operator ${\cal A}$ can be read from
Eq.~\eqref{matrix-equations}. 
For the static background (\ref{stat2}) the operator ${\cal A}$ is
non-degenerate. Thus, it is non-degenerate for non-trivial
backgrounds, which are close enough to (\ref{stat2}). One concludes
that the matrix $\hat s$ vanishes,
\be
\label{symm}
\d_i\beta^b+\d_b\beta^i=0\;.
\ee
Differentiating Eq.~(\ref{symm}) twice yields the Laplace equation for
the divergence, $\Delta \d_a\beta^a=0$. Imposing vanishing boundary
conditions at spatial infinity we obtain $\d_a\beta^a=0$. Then,
differentiating Eq.~(\ref{symm}) once, we find that all the
components $\beta^a$ vanish.

Thus the longitudinal modes $\phi_8$ and $\psi_{6\alpha}$ 
do not become propagating in 
non-trivial background. This result implies the absence of 
ghosts and/or tachyons in the model \eqref{action} to higher orders in
classical perturbation theory.


\end{document}